\documentclass[final,a4paper,5p,times,twocolumn]{article}

 \usepackage{graphicx}

\usepackage{amssymb}

\usepackage{lineno}

\usepackage{subfigure}
\usepackage{caption}
\usepackage{siunitx}
\usepackage[ansinew]{inputenc}

\title{Parametrization of the radiation induced leakage current increase of NMOS
transistors}

\author{M. Backhaus\\
\emph{CERN} \\ \\
Now: \emph{ETH Z\"urich, Otto-Stern-Weg 5, 8093 Z\"urich, Switzerland}\\
 E-mail: \emph{backhaus@cern.ch}}

\date{\today}

\begin{document}

\twocolumn[
  \begin{@twocolumnfalse}

\maketitle

\begin{abstract}
The increase of the leakage current of NMOS transistors during exposure to
ionizing radiation is known and well studied. Radiation hardness by
design techniques have been developed to mitigate this effect and
have been successfully used. More recent developments in smaller feature size
technologies do not make use of these techniques due to their drawbacks in terms
of logic density and requirement of dedicated libraries. During
operation the resulting increase of the supply current is a serious challenge
and needs to be considered during the system design.\\

A simple parametrization of the leakage current of NMOS transistors as a
function of total ionizing dose is presented. The parametrization uses a transistor
transfer characteristics of the parasitic transistor along the shallow trench
isolation to describe the leakage current of the nominal transistor. Together
with a parametrization of the number of positive charges trapped in the silicon
dioxide and number of activated interface traps in the silicon to silicon
dioxide interface the leakage current as a function of the exposure time to
ionizing radiation results. This function is fitted to data of the leakage
current of single transistors as well as to data of the supply current of full ASICs.\\\\\\\\\\
\end{abstract}


\end{@twocolumnfalse}
]

\section{Introduction}
The radiation induced leakage current of NMOS transistors and the
threshold voltage shift are well known and intensively studied challenges for
the design of radiation hard application specific integrated circuits (ASIC).
Hardness by design (HBD) techniques \cite{hbd1}\cite{hbd2} have been developed
and extensively used in the circuit design in the \SI{0.5}{\micro\meter} technology node for the LHC 
experiments \cite{faccio_lhc}. The HBD techniques consist mainly of the use of
enclosed gate transistors to mitigate the source to drain leakage current along the shallow trench
isolation (STI) and guard ring structures surrounding the transistors to avoid
leakage current between neighboring structures. The use of this  technique
requires an increased area as well as custom libraries. With the transition to the
\SI{0.13}{\micro\meter} technology node, the amplitude of the leakage current
increase decreases by three orders of magnitude, with the result that the HBD
techniques are needed in sensitive nodes of the design only for radiation hard
designs \cite{faccio_130}. However, while the chip is operational, the increase
of the leakage current of NMOS transistors results in a significant increase of the supply
current. A generic parametrization of the leakage current as a
function of total ionizing dose (TID) of single NMOS transistors is presented 
in this paper. The resulting function is fitted to published data of a diversity
of technologies in use or under investigation for use in extremely radiation
intense environments. As demonstrated in this paper, this parametrization can be
used as well to model the supply current shift with TID on full ASICs and to
predict the current to be expected during operation of the ASICs as a function
of the temperature and the dose rate, once the parameters of the parametrization
are measured.

\section{Parametrization of the leakage current}
\label{sec:parametrization}
\begin{figure}[htbp]
\centering
\includegraphics[width=0.5\linewidth]{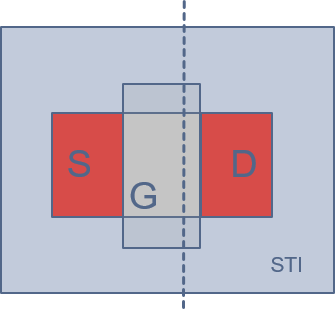}
\caption{Top View of a linear transistor. The source (S), gate (G), and drain
(D) as well as the shallow trench isolation (STI) are indicated. The dashed
line shows the place of the cross section in figure \ref{fig:CrossSection1}.}
\label{fig:TopView}
\end{figure}

This model describes the leakage current increase of linear NMOS transistors
independent of technology details. The increase of the leakage current as a 
function of total ionizing dose has been reported for a large number of
technologies.
It originates from \emph{parasitic transistor channels} along the STI, in which the 
transistor is embedded as indicated in figure \ref{fig:TopView}.\\

\begin{figure}[htbp]
\centering
\includegraphics[width=0.5\linewidth]{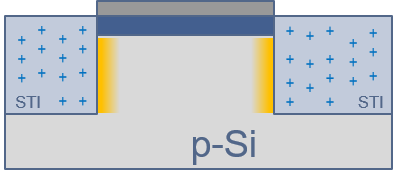}
\caption{Cross section of a transistor along the line indicated in figure
\ref{fig:TopView}. The channel of the parasitic transistor along the STI is
indicated in yellow.}
\label{fig:CrossSection1}
\end{figure}

With a positive space charge appearing in the STI due to ionizing radiation, an
inversion layer is created along the STI, which results in a leakage current
paths from source to drain \cite{faccio_130}. This is indicated in the cross
section in figure \ref{fig:CrossSection1}.\\

\begin{figure}[htbp]
\centering
\includegraphics[width=0.5\linewidth]{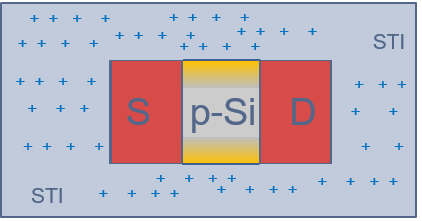}
\caption{View from the top in the transistor transistor below the gate. The
channel of the parasitic transistor along the STI is indicated in yellow.}
\label{fig:CrossSection2}
\end{figure}

This leakage current paths can be described as the channel of a parasitic
transistor. As can be seen in figure \ref{fig:CrossSection2}, the parasitic
transistor has a layout analogue to a linear transistor with the only difference,
that the electric field opening the transistor channel originates from the
positive space charge in the STI instead of from the potential at the gate.
The space charge in the STI is always positive and therefore in PMOS 
transistors this transistor channel is closed by the radiation induced 
space charge. Thus the leakage current increase is only observed in NMOS
transistors.\\

This analogue layout motivates to describe the leakage current increase due to
the parasitic transistor using the \emph{transfer characteristics} of
the parasitic transistor.

\subsection{Transfer characteristics of the parasitic transistor}
\label{sec:sub:transfer}
The transfer characteristics (drain current as a function of the gate to source
voltage) for transistors operated in saturation mode can be simplified by 
\begin{eqnarray}
	I_D &\approx& 0
	\;\;\;\;\;\;\;\;\;\;\;\;\;\;\;\;\;\;\;\;\;\;\;\;\;\;\;\;\mbox{for} \;\;
	V_G < V_{thr}
	\nonumber
	\\
	I_D &\approx& K' \cdot (V_G - V_{thr})^2 \;\;\;\mbox{for} \;\; V_G \geq
	V_{thr}
\end{eqnarray}
where sub-threshold leakage is neglected. $I_D$ is the drain to source current,
$V_G$ the gate to source voltage, and $V_{thr}$ the threshold voltage. $K'$ is
the proportionality constant containing the widths, lengths, oxide
capacitance, and the mobility of the minority charge carriers in the channel of
the transistor, etc. In the parasitic transistor (which is responsible for the
leakage current) the electric field originates from the effective space charge. Assuming
an electric field in the silicon proportional to the effective space charge, the
gate voltage of the parasitic transistor is in this model replaced by the
\emph{effective number of charges} $N_{eff}$. Similarly, the threshold voltage
is expressed by the \emph{threshold number of charges} $N_{thr}$.
The transfer characteristics of the parasitic transistor are then given by 
\begin{eqnarray}
	I_{D}^{par} &=& 0 \nonumber
	\;\;\;\;\;\;\;\;\;\;\;\;\;\;\;\;\;\;\;\;\;\;\;\;\;\;\;\;\;\;\mbox{for} \;\; N_{eff} <
	N_{thr}\nonumber \\
	 I_{D}^{par} &=& K \cdot (N_{eff} - N_{thr})^2
	\;\;\mbox{for} \;\; N_{eff} \geq N_{thr} \nonumber
	\label{eqn:transferChar}
\end{eqnarray}
and thus the leakage current of the regular transistor $I_{leak}$ is given
by
\begin{eqnarray}
	I_{leak} &=& I_{leak}^0 \nonumber \\
	&& \mbox{for} \;\; N_{eff} < N_{thr}\nonumber \\
	\nonumber \\
	 I_{leak} &=& I_{leak}^0 + K \cdot (N_{eff} - N_{thr})^2 \nonumber \\
	&& \mbox{for} \;\; N_{eff} \geq N_{thr}
	\label{eqn:transferChar}
\end{eqnarray}
with $I_{leak}^0$ the preirradiation leakage current of the transistor.
$N_{thr}$ is constant in time (and therefore accumulated dose) and temperature.
With these assumptions a parametrization of the number of effective charge carriers $N_{eff}$
describes the leakage current shift in NMOS transistors.

\subsection{Processes of charge generation}
A combination of four processes results in the effective space charge in the
STI.
First, due to ionization of the atoms, the incident radiation generates free
electron hole pairs in the silicon dioxide of the STI
\cite{ehpairs1}\cite{ehpairs2}. Some of these pairs recombine quickly, but the
mobility of the electrons is between six and twelve orders of magnitude larger than the mobility of the holes, depending on the
temperature and the electrical field [7-10].
Therefore many electrons are quickly removed from the STI, while the left over holes move slowly in the silicon
dioxide by hopping transport \cite{ehpairs1}. Sites with missing oxide atoms
in the amorphous silicon dioxide result in energy levels above the valence band and
thus electrically neutral deep hole traps \cite{ehpairs1}. These traps are
distributed in the STI volume, and their concentration is largly influenced by the manufacturing of
the STI, and thus it is technology dependent. During the movement some holes get
trapped in these sites and a space charge is built. The holes have a certain
probability to get detrapped by thermal energy. The lifetime $\tau_{ox}$ of the
holes in the traps depends on the energy level of the traps, and of the
temperature. \\

Holes which are not trapped in the silicon dioxide can move to the silicon to
silicon dioxide interface. At this interface are incomplete or dangling atomic
bonds due to the abrupt transistion from amorphous to christalline material.
The dangling bonds manifest themselves as energy levels in the
band-gap, and thus they trap mainly electrons or holes, depending on the
Fermi-level of the silicon \cite{ehpairs1}. This trapping of electrons in the
case of NMOS transistors (p-type silicon) and holes in the case of PMOS transistors
(n-type silicon) degrades the transistor performance, and therefore usually the
dangling bonds are deactivated by the manufacturer using hydrogen. The
radiation induced free holes can react with the hydrogen and the dangling bonds
get re-activated \cite{ehpairs1}. In this case they are commonly called radiation induced
\emph{interface traps}.\\

In the case of NMOS transistors the activation of the interface traps results in
a negative space charge, while the trapping of holes in the STI results in a
positive charge \cite{faccio_130}. The electric fields compensate each other, so
that the effective number of charges $N_{eff}$:
\begin{eqnarray}
	N_{eff} &=& N_{ox} - N_{if}
\end{eqnarray}
becomes relevant, with $N_{ox}$ the number of trapped holes in the STI, and
$N_{if}$ the number of electrons trapped at the interface. The parametrization
of these two numbers during and after ionizing radiation is explained in the
following.

\subsubsection{Parametrization of the number of positive charges
trapped in the STI}
During exposure to ionizing radiation with a constant dose rate $D$
the number of holes getting trapped in the STI volume is proportional to the
exposure time $t$ with a proportionality constant $k_{ox} D$, where $k_{ox}$
describes how many holes are trapped per dose unit. At the same time, the
trapped holes have a life-time $\tau_{ox}$ in the traps, until they are free again
and can move out of the STI. Therefore, the number of holes trapped in
the STI is defined by the differential equation 
\begin{eqnarray}
\frac{d}{dt} N_{ox}(t) &=& k_{ox} D \; - \; \frac{1}{\tau_{ox}} N_{ox}(t)
\label{eqn:differential}  
\end{eqnarray}
which is solved by 
\begin{eqnarray}
N_{ox}(t) &=& k_{ox} D \cdot \tau_{ox} \cdot \left(1 - e^{-\frac{t}{\tau_{ox}}}\right).
\end{eqnarray}
If the irradiation stops after an exposure time $t_1$, only the
second term of equation (\ref{eqn:differential}) stays and results in
and exponential decrease of the number of holes trapped in the STI, which is
usually called annealing. This decrease can be described by 
\begin{eqnarray}
N_{ox}(t) &=& N_{trap} \nonumber \\
&&+ \left(N_{ox}(t_1) - N_{trap}\right) \cdot
e^{-\frac{t-t_1}{\tau_{ox}}}
\label{eqn:annealing}
\end{eqnarray}
where $N_{trap}$ describes the number of holes captured in traps of the oxide
which are too deep for detrapping at the given temperature.

\subsubsection{Parametrization of the number of activated interface traps}
The number of activated interface traps follows a very similar behavior, as
motivated here.  The holes travelling to the silicon to silicon dioxide
interface and activating  the radiation induced interface traps are generated
again with a constant rate $k_{if}D$. $k_{if}$ describes  here the  number of
holes available to activate interface traps per dose unit. The number  of 
interface traps that can be activated by the radiation is technology dependent
and it is limited. Therefore the probability that the holes activate new
interface  traps decreases with time exponentially. This can be described
equivalently using the analogue equation
\begin{eqnarray}
N_{if}(t) &=& k_{if} D \cdot \tau_{if} \cdot \left(1 -
e^{-\frac{t}{\tau_{if}}}\right).
\end{eqnarray}

The needed temperature to anneal the interface traps is known to be well above
room temperature and even more above the operational temperature of the ASICs.
It is technology dependent and in the range of \SI{100}{\celsius} to
\SI{300}{\celsius}. Therefore, the annealing of the interface traps is
negligible for the presented parametrization. 

\subsection{Summary of the parametrization}
\label{sec:sub:parasum}
The complete formula for the leakage current during irradiation is therefore
given by
\begin{eqnarray}
	I_{leak} &=& I_{leak}^0
	\label{eqn:IleakFullZero}
\end{eqnarray}
for
\begin{eqnarray}
	k_{ox} D \cdot \tau_{ox} \cdot \big(1 - e^{-\frac{t}{\tau_{ox}}}\big)
	\label{eqn:ileakfull0} \\
	- k_{if} D \cdot \tau_{if} \cdot \big(1 - e^{-\frac{t}{\tau_{if}}}\big) &<&
	N_{thr} \nonumber
\end{eqnarray}	
and by 
\begin{eqnarray}
	 I_{leak} &=& I_{leak}^0 \nonumber \\
	 &&+ K \cdot \Big[k_{ox} D \cdot \tau_{ox} \cdot \big(1 -
	 e^{-\frac{t}{\tau_{ox}}}\big) \nonumber \\
	&&- k_{if} D \cdot \tau_{if} \cdot \big(1 - e^{-\frac{t}{\tau_{if}}}\big) \nonumber \\
	&&- N_{thr}\Big]^2
	\label{eqn:ileakfull1}
\end{eqnarray}
for
\begin{eqnarray}
	k_{ox} D \cdot \tau_{ox} \cdot \big(1 - e^{-\frac{t}{\tau_{ox}}}\big)
	\nonumber \\ 
	- k_{if} D \cdot \tau_{if} \cdot \big(1 - e^{-\frac{t}{\tau_{if}}}\big) &\geq&
	N_{thr}. \nonumber
\end{eqnarray}	

During the periods with no incident ionizing radiation (dose rate $D = 0$)
$N_{ox}$ is similarly replaced by equation (\ref{eqn:annealing}), and the number
of activated interface traps $N_{if}$ stays constant because they do not anneal
at the considered temperature range. 
\begin{figure}[htbp]
\centering
		\includegraphics[width=\linewidth]{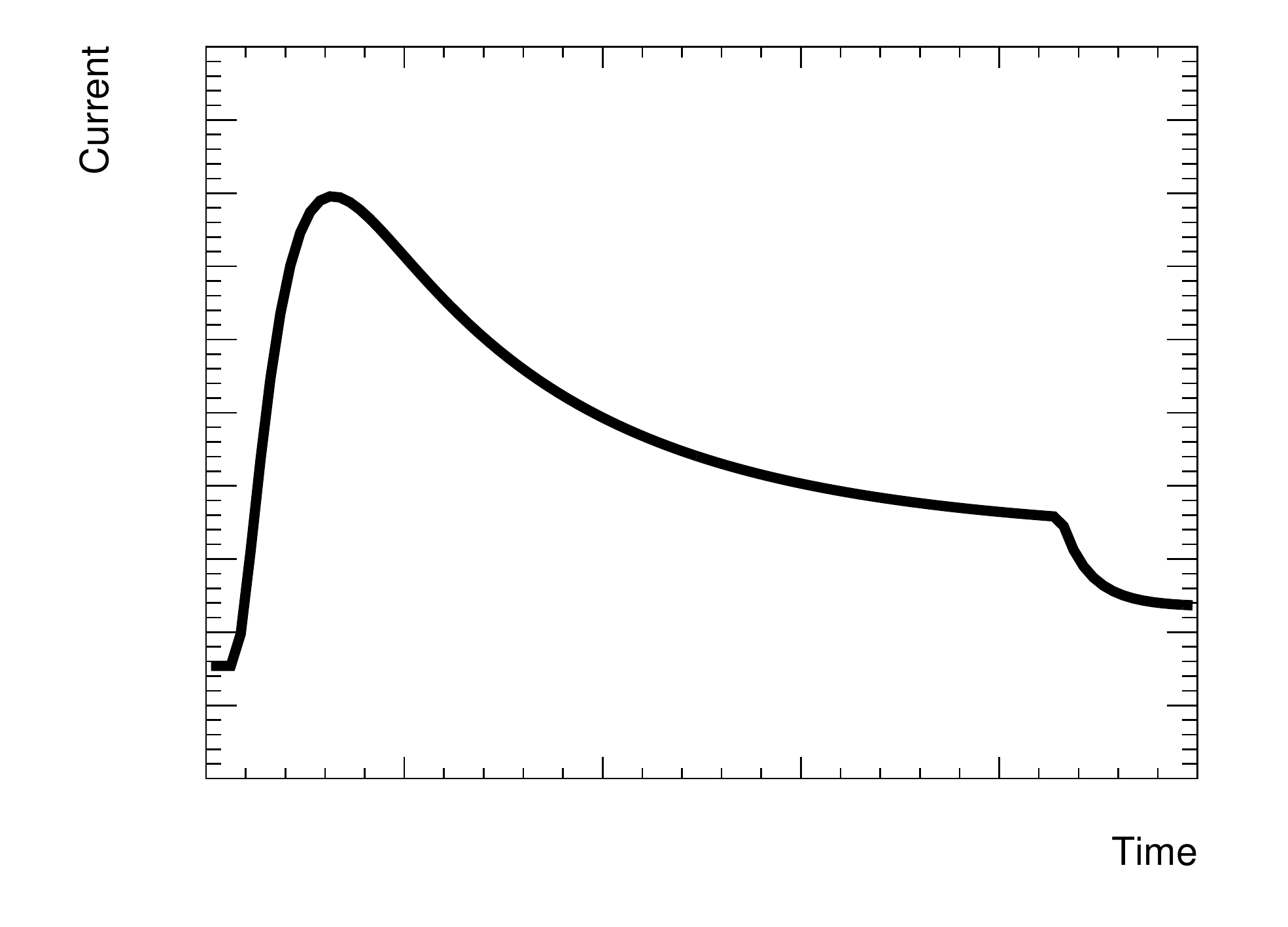}
\caption{Plot of the leakage current parametrization as a function of
the time, including the abrupt switch-off of the ionizing radiation, resulting
in the exponential decrease.}
\label{fig:IleakExample}
\end{figure}

Figure \ref{fig:IleakExample} presents the
resulting leakage current increase as a function of time during an exposure to
ionizing radiation with a constant dose rate. The exponential decrease step in 
the far right of the plot illustrates the annealing behavior when the radiation 
is switched-off. For many studies the leakage current is given as a function of 
the TID. The parametrization can be expressed during periods of constant 
instensity exposure using $\mbox{TID} = D \cdot t$.\\

The temperature dependency of the parameters is not explicitely included here. 
The generation of the positive space charge depends on the temperature. This can 
be modelled sufficiently well with a temperature dependent de-trapping probability, 
and neglecting the temperature dependence of the electron and hole mobility and 
generation \cite{hmobility1}\cite{generation_yield}. Then 
only $\tau_{ox}$ directly depends on the temperature. The generation of the interface traps 
is also expected to be a function of the temperature, but it is not well known. A 
dedicated measurement campaign is ongoing to provide the data that are needed 
for the extraction of the temperature dependence of these terms. As these data 
are not yet available, here the parametrization is used to describe the measured 
leakage current increase at a given temperature.\\

In the following two sections this parametrization is first fit to published
data of single NMOS transistors. Then the same function is used to describe
the supply current shift of the ATLAS IBL pixel readout chip \mbox{FE-I4}
\cite{fe-i4} as an example of the consequence of the leakage current shift for the ASIC
operation in radiation intense environments. This is an example for the power of
this parametrization to predict the ASIC supply current increase once the basic
parameters are known for the given technology.

\section{Fit to single transistor data}
\label{sec:singleTransistorFits}
\begin{figure}[htbp]
\centering
		\includegraphics[width=\linewidth]{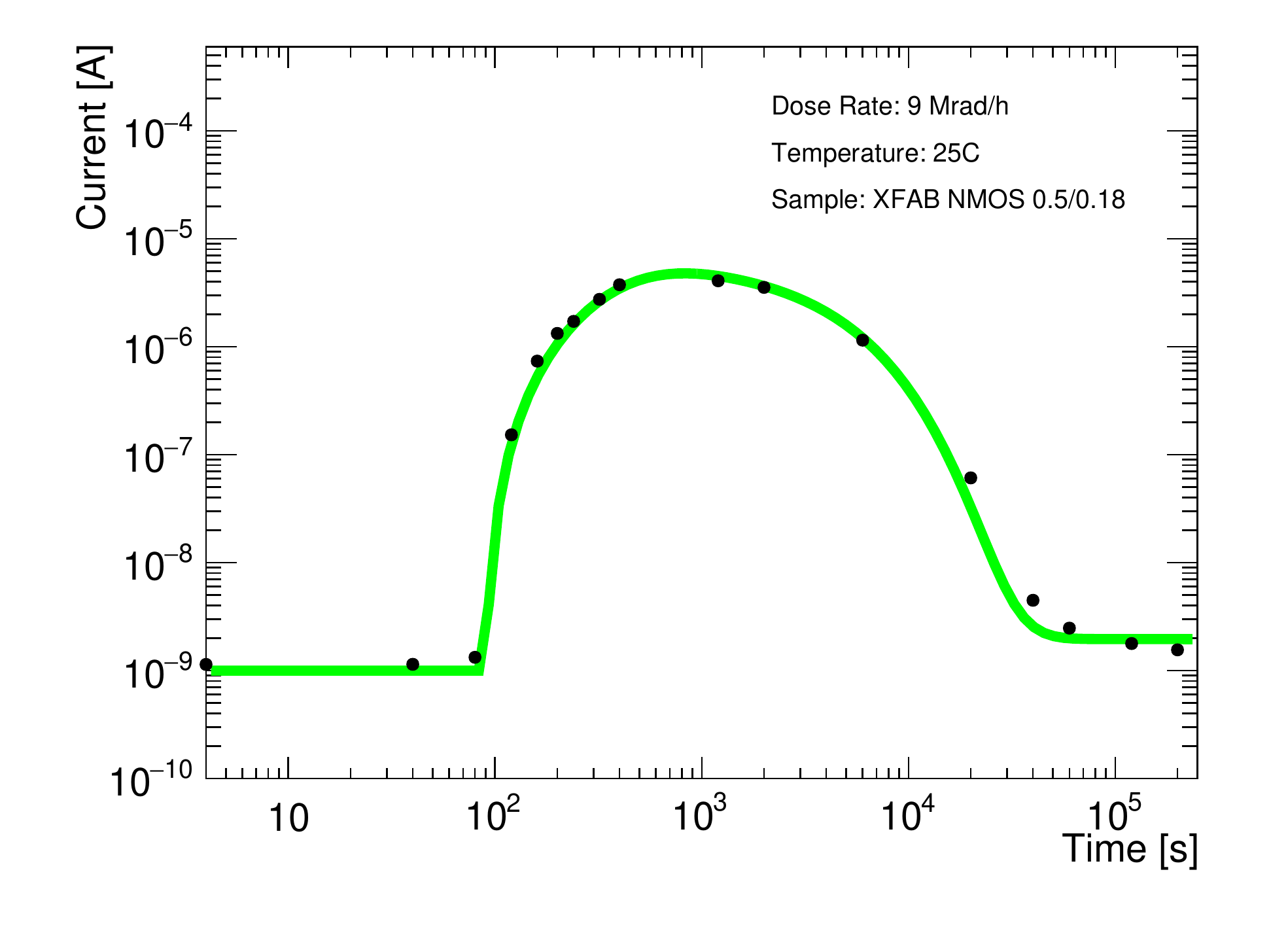}
\caption{Fit of the leakage current parametrization as a function of
the time to the leakage current measurement of a single NMOS transistor
(width \SI{0.5}{\micro\meter}, length \SI{0.18}{\micro\meter}). The
data were previously published in \cite{xfab}.}
\label{fig:XFAB_n1}
\end{figure}

\begin{figure}[htbp]
\centering
		\includegraphics[width=\linewidth]{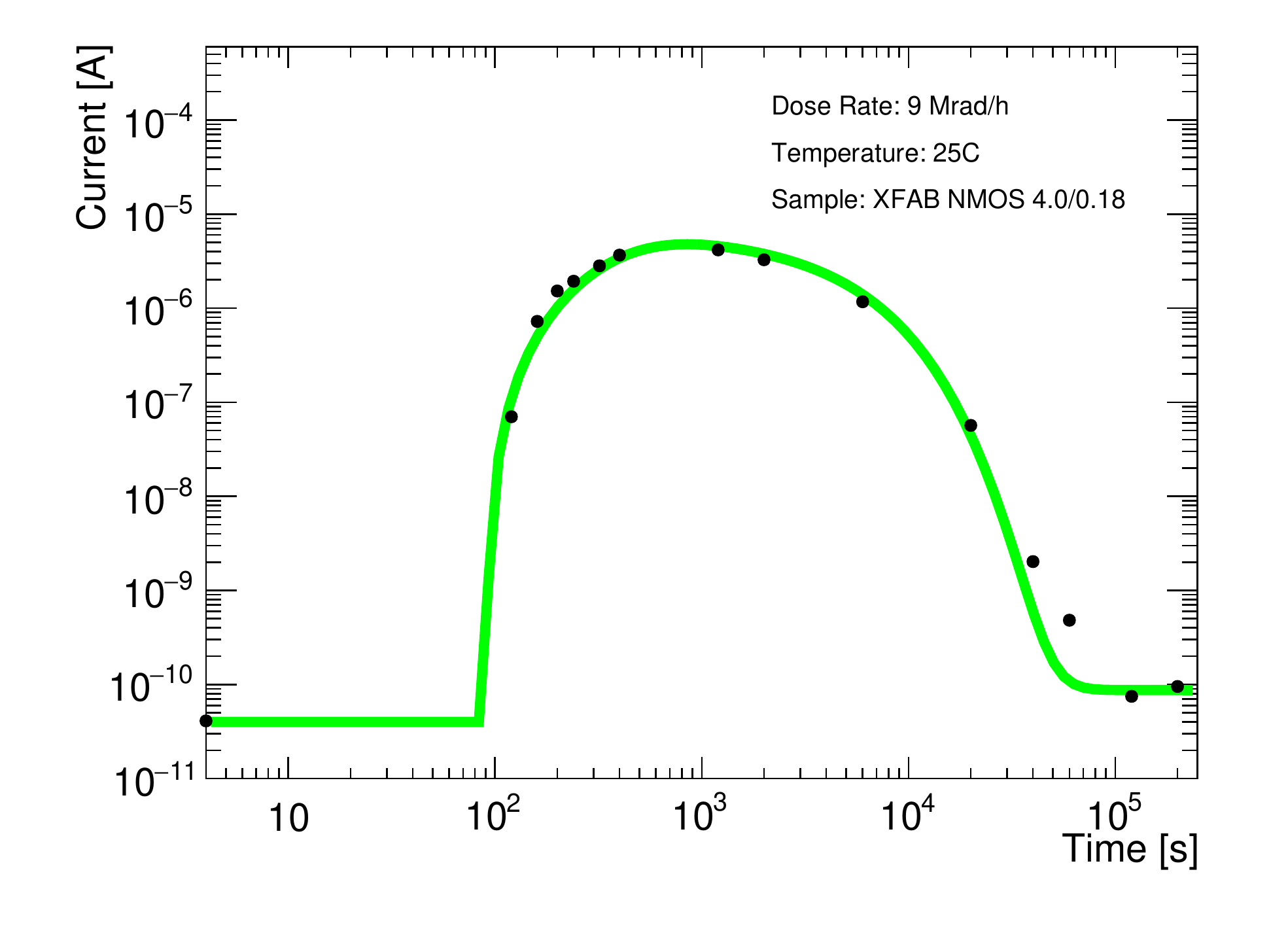}
\caption{Fit of the leakage current parametrization as a function of
the time to the leakage current measurement of a single NMOS transistor
(width \SI{2.0}{\micro\meter}, length \SI{1.4}{\micro\meter}). The
data were previously published in \cite{xfab}.}
\label{fig:XFAB_n3}
\end{figure}

\begin{figure}[htbp]
\centering
		\includegraphics[width=\linewidth]{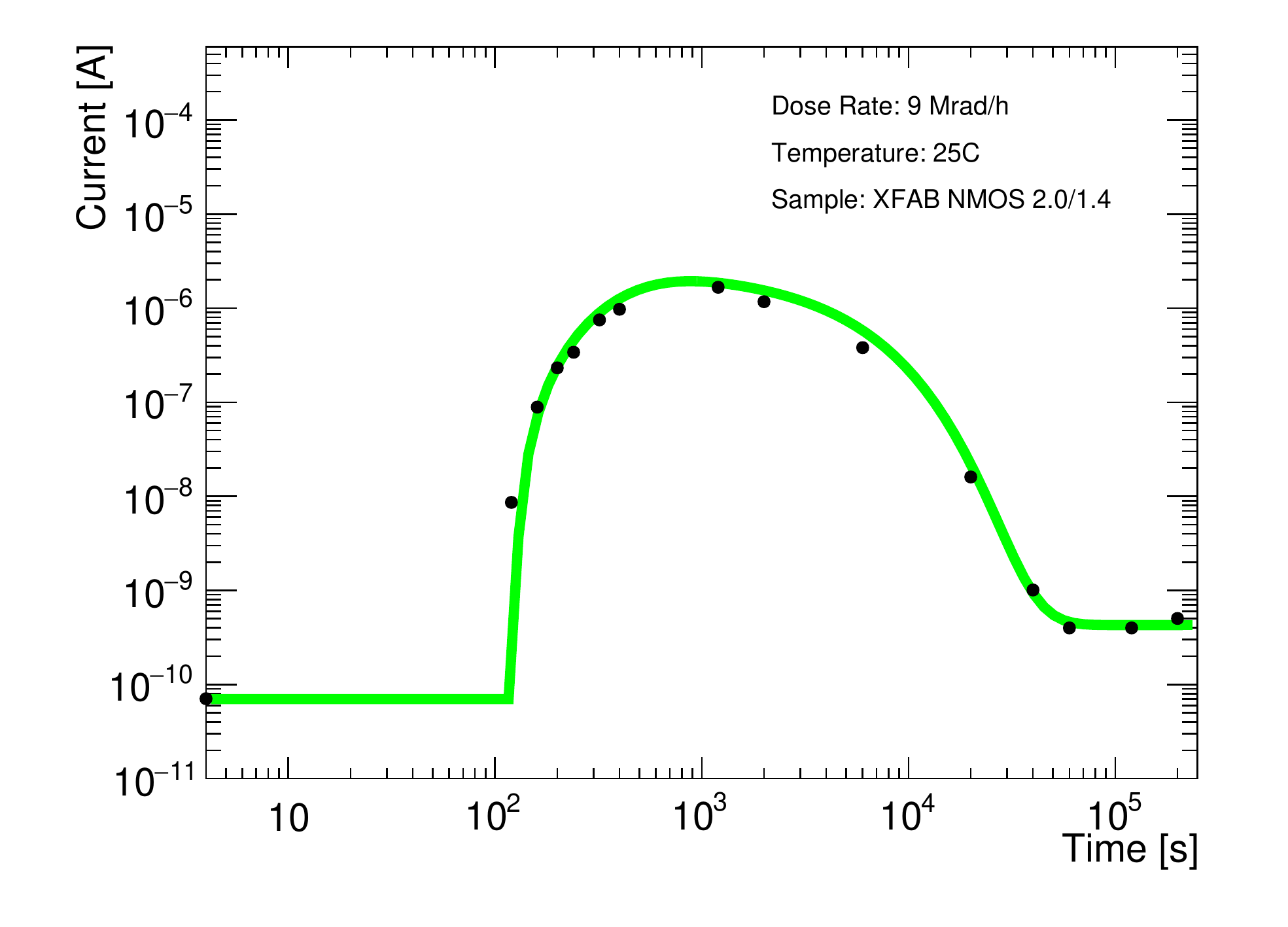}
\caption{Fit of the leakage current parametrization as a function of
the time to the leakage current measurement of a single NMOS transistor
(width \SI{4.0}{\micro\meter}, length \SI{0.18}{\micro\meter}). The
data were previously published in \cite{xfab}.}
\label{fig:XFAB_n14}
\end{figure}

Equation (\ref{eqn:IleakFullZero}), (\ref{eqn:ileakfull0}) and
(\ref{eqn:ileakfull1}) are used to describe the leakage current shift of single
NMOS transistors produced in a 180nm silicon on insulator (SOI) process \cite{xfab_process}. The data have been
published previously in \cite{xfab}, and were obtained at a dose  rate $D$ of
\SI{9}{\mega\radian\per\hour} at a temperature of \SI{\sim25}{\celsius}.
Because the proportionality factor $K$ appears in the parametrization only in products
with the other parameters, $K$ was fixed to a value of \SI{1e-19}{\ampere} per
effective charge carrier. The threshold charge $N_{thr}$, the time constants
$\tau_{ox}$ and $\tau_{if}$, as well as the proportionality constants $k_{ox}$
and $k_{if}$ are free fit parameters. The data and the function are shown in
figure \ref{fig:XFAB_n1} to \ref{fig:XFAB_n14} and demonstrate the good
agreement of data and parametrization on single transistor level.


\section{Fit to full ASIC supply current shift}
ASICs composed of a large number of linear NMOS transistors can show a
significant supply current shift when operated under ionizing radiation. This
supply current shift is a serious challenge for the ASIC operation and impacts
the design of the system, because the services need to be able to cope with
this shift. The amplitude of the increase depends on the environmental
conditions, such as dose rate and temperature. An intense investigation program
is currently carried out on ATLAS \mbox{FE-I4} readout chips. The ASIC, produced
in IBM \SI{130}{\nano\meter} technology, is composed of about 80 million
transistors, and HBD techniques are not used for the large majority of the
transisors. The ASIC is operated under controlled environmental conditions while
being exposed to x-ray radiation. The supply current is measured as a
function of the exposure time. This measurement is carried out for various dose
rates and temperatures. Some preliminary results are public \cite{fe-i4-data}
and used here to demonstrate the ability of the parametrization to describe the supply current
shift of full ASICs.\\

\begin{figure}[htbp]
\centering
		\includegraphics[width=\linewidth]{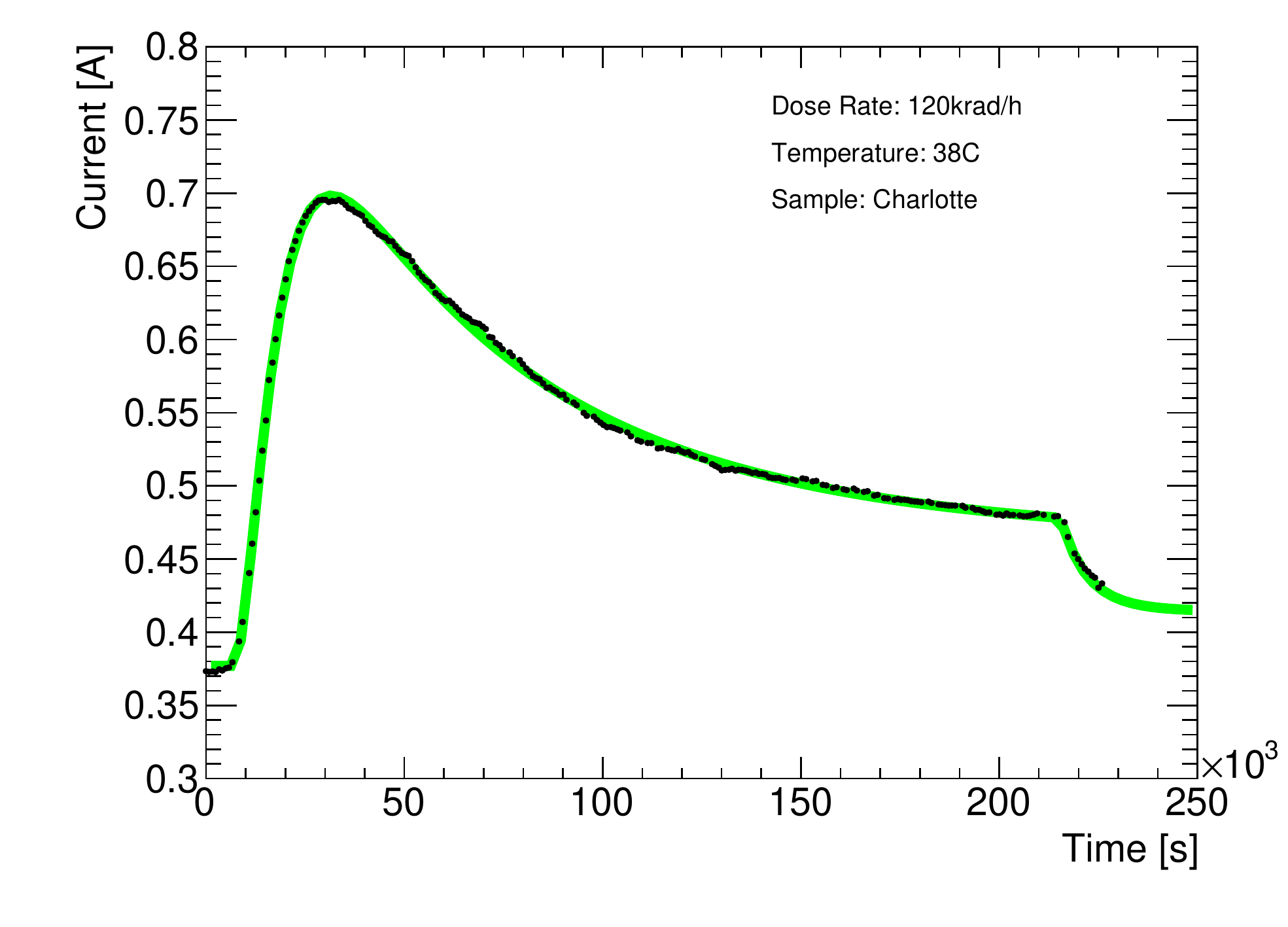}
\caption{Fit of the leakage current parametrization as a function of
the time to the supply current measurement of an ATLAS \mbox{FE-I4}
readout-chip.
At a the time $t_1 = \SI{215500}{\second}$ the dose rate is switched-off, and the
annealing starts. The data were previously published in \cite{fe-i4-data}.}
\label{fig:Charlotte}
\end{figure}

Figure \ref{fig:Charlotte} shows the fit to the supply current of the ASIC as a
function of the exposure time using \SI{120}{\kilo\radian\per\hour} dose rate at
\SI{38}{\celsius}. The same parameters are as in section
\ref{sec:singleTransistorFits} free during the fit, while $K$ is fixed now to
\SI{1e12}{\ampere} per effective charge in order to account for that the supply current is the
convolution of the leakage current of about several ten millions of transistors.
Additionally, the pre-irradiation supply current is added as offset. At the
time $t_1 = \SI{215500}{\second}$ the irradiation was switched-off and the
annealing as described in equation (\ref{eqn:annealing}) is shown. The
parameters are fit to the time interval 0 to $t_1$ only. For the annealing the
same parameters are used, especially the same time constant $\tau_{ox}$. This
reflects that the annealing is caused be the same process as the saturation of
the number of positive charges trapped in the oxide during irradiation.\\

\begin{figure}[htbp]
\centering
		\includegraphics[width=\linewidth]{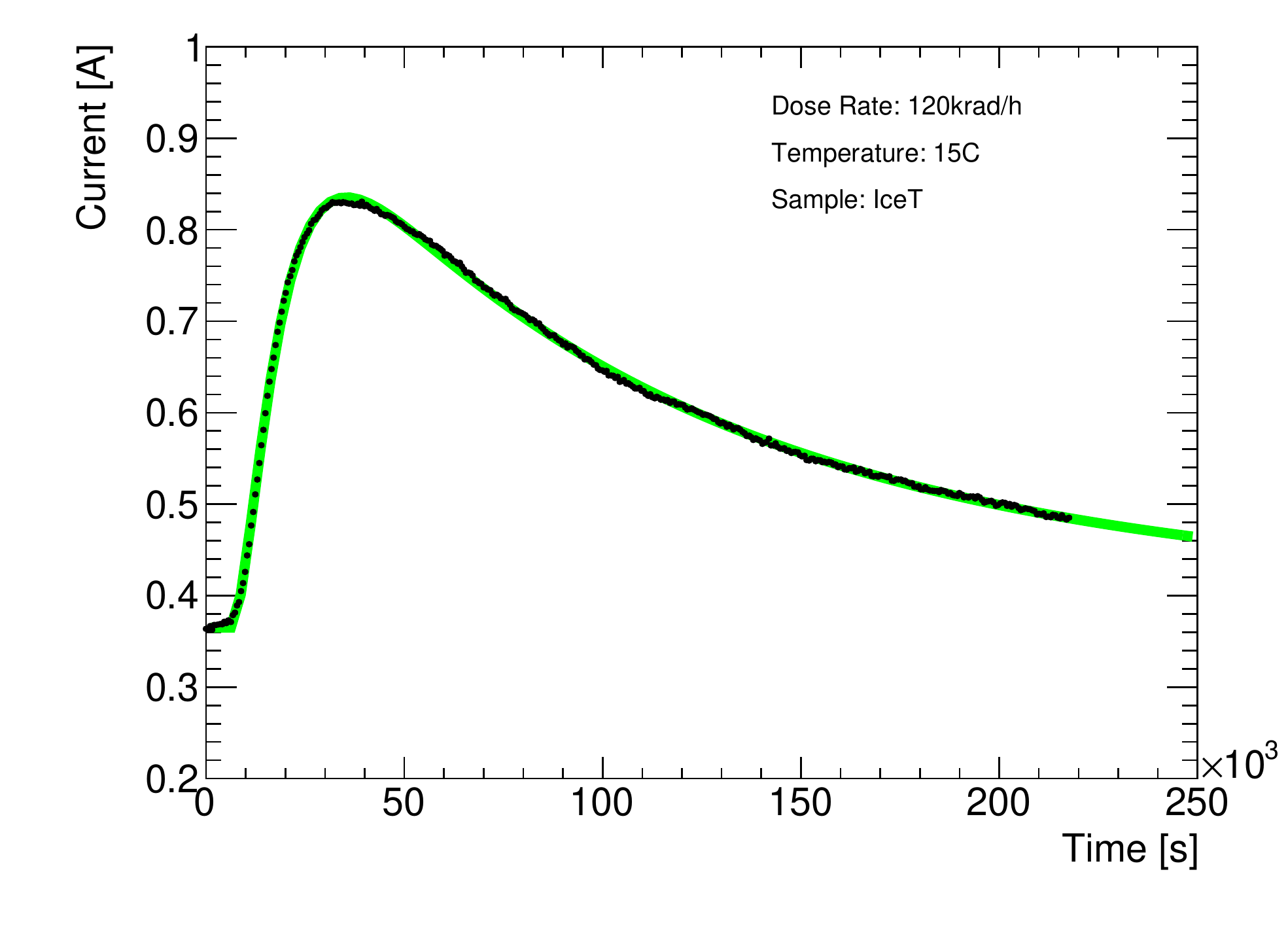}
\caption{Fit of the leakage current parametrization as a function of
the time to the supply current measurement of an ATLAS \mbox{FE-I4}
readout-chip.
The data were previously published in \cite{fe-i4-data}.}
\label{fig:IceT}
\end{figure}

\begin{figure}[htbp]
\centering
		\includegraphics[width=\linewidth]{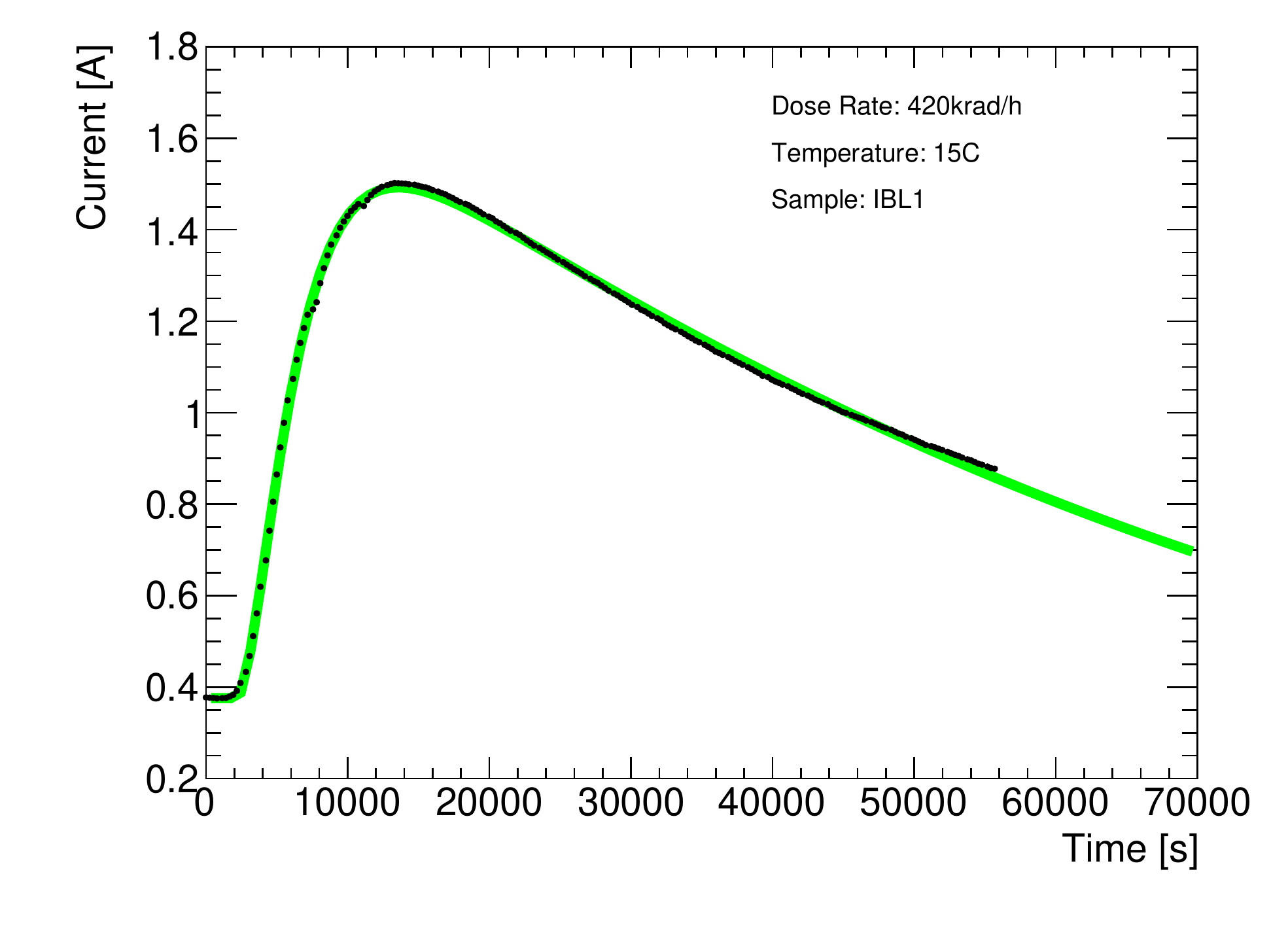}
\caption{Fit of the leakage current parametrization as a function of
the time to the supply current measurement of an ATLAS \mbox{FE-I4}
readout-chip.
The data were previously published in \cite{fe-i4-data}.}
\label{fig:IBL1}
\end{figure}

Figure \ref{fig:IceT} shows the same fit to data taken at
different temperature (\SI{15}{\celsius}). A slightly higher amplitude of
the increase is observed, as expected due to the longer lifetime $\tau_{ox}$ of
the positive charges in the traps. 
The data shown in figure \ref{fig:IBL1} are taken again at \SI{15}{\celsius},
but using a higher dose rate of \SI{420}{\kilo\radian\per\hour}. The
current maximum is significaltly higher due to the larger dose rate.\\ 

The exact dependencies of the fit parameters from the temperature and from
the dose rate are currently under investigation. The good agreement between the
parametrization and the data demonstrate the power of this parametrization to predict the
supply current during operation in radiation environment, once the basic
parameters have been measured for the ASIC in the laboratory.

\section{Conclusions}
The presented parametrization of the leakage current of NMOS transistors uses
simplified transfer characteristics of the \emph{parasitic} source to drain
transistor along the STI. The good agreement of single transistor
measurement data and the resulting function show that the description
of the gate potential and threshold voltage by the effective number of charges
and threshold charge, as well as the description of their concentration as a
function of the exposure time describes the observations.\\

Furthermore, the parametrization can directly be used to predict the supply
current increase of full ASICs in radiation environment, when HBD techniques are
not used for the majority of transistors, once the parameters are known as a
function of dose rate and temperature.

\section*{Acknowledgements}
The author would like to thank S. Fernandez-Perez for the agreement to use the
single transistor data for the presented study. A similar thanks belongs to A.
LaRosa (MPI Munich), K. Dette (CERN), T. Obermann (University of Bonn) and D.
Sultan (INFN Trento) for the collaboration in the \mbox{FE-I4} irradiations.\\
A special thanks goes to F. Faccio (CERN) for the always fruitful discussions
and helpful explanations, as well as for proof-reading each draft of the work.

\end{document}